\documentclass[a4paper,11pt]{amsart}
\usepackage[utf8]{inputenc}
\usepackage{amsfonts,amssymb,amsmath,amscd, amsthm, xcolor}
\usepackage{graphicx}
\usepackage{caption, subcaption,float}
\usepackage{url}
\usepackage[ruled, linesnumberedhidden,vlined]{algorithm2e} 

\title{The root extraction problem in braid group-based cryptography}
\author[M. Cumplido]{María Cumplido}
\address{María Cumplido: Departmento de Álgebra,
Facultad de Matemáticas, Universidad de Sevilla, Spain}
\email{cumplido@us.es}
\author[D. Kahrobaei]{Delaram Kahrobaei}
\address{Delaram Kahrobaei: The City University of New York, Queens College and CUNY Graduate Center, U.S.A., University of York, U.K.}
\email{dkahrobaei@gc.cuny.edu}
\author[M. Noce]{Marialaura Noce}
\address{Marialaura Noce: Dipartimento di Matematica, Università degli Studi di Salerno, Italy}
\email{mnoce@unisa.it}

\newcommand{\po}{\preccurlyeq}


\begin{document}

\maketitle
\begin{abstract}
The root extraction problem in braid groups is the following: given a braid $\beta \in \mathcal{B}_n$ and a number $k\in \mathbb{N}$, find $\alpha\in \mathcal{B}_n$ such that $\alpha^k=\beta$. In the last decades, many cryptosystems  such as authentication schemes and digital signatures  based on the root extraction problem have been proposed. In this paper, we first describe these cryptosystems built around braid groups. Then we prove that, in general, these authentication schemes and digital signature are not secure by presenting for each of them a possible attack.
\end{abstract}

\section{Introduction}

In the past decades, cryptographers have focused their attention in braid groups to use them as a tool to construct public key exchanges \cite{AAG, KoLee}, authentication schemes and digital signatures. It is interesting to mention about the recent work of Anshel, Atkins, Goldfeld, and Gunnells on the WalnutDSA$^{TM}$ digital signature \cite{AaGG} which has been claimed by its authors to be quantum resistant, which uses braid groups as a platform. In this paper, we will do an exhaustive analysis of proposed cryptosystems based on the root extraction problem for braid groups. 

\begin{figure}
    \centering
    \includegraphics[width=0.8\linewidth]{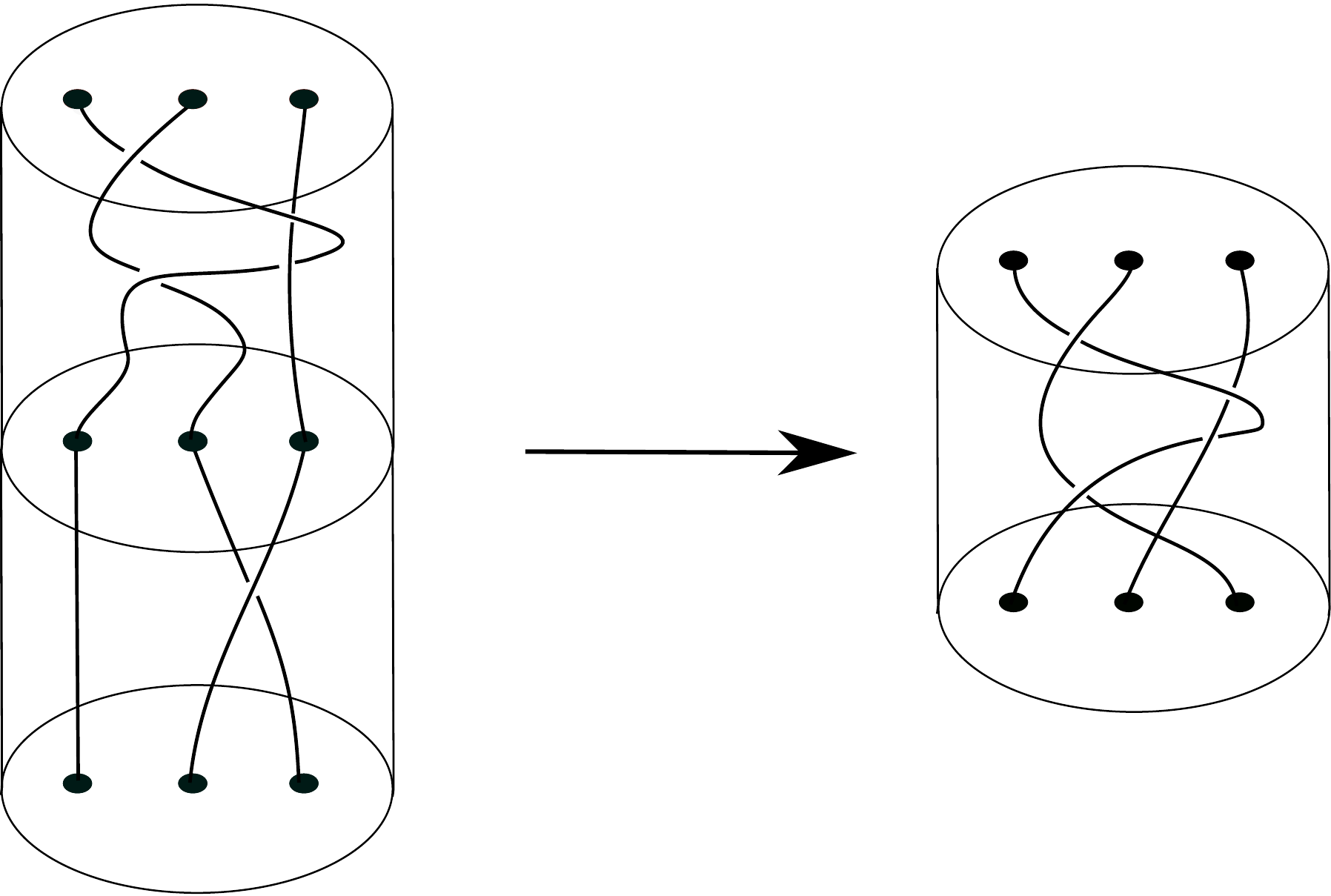}
    \caption{How to multiply the braids $\sigma_1\sigma_2^{-2}\sigma_1$ and $\sigma_2$.  }
    \label{fig:multi}
\end{figure}

Braid groups were first defined by Artin in 1947 \cite{Artin}. A braid group  $\mathcal{B}_n$ on $n$ strands is the group with the following presentation:

$$\mathcal{B}_n= \left\langle \sigma_1,\sigma_2, \dots, \sigma_{n-1} \, \left| \, \begin{array}{cc}
     \sigma_i\sigma_j\sigma_i=\sigma_j\sigma_i\sigma_j, & |i-j|=1  \\
         \sigma_i\sigma_j=\sigma_j\sigma_i, & |i-j|>1  
     \end{array}\right.
 \right\rangle. $$
Graphically a braid on $n$ strands can be seen as a collection of $n$ paths in a cylinder joining $n$ distinguished points at the top of the cylinder with $n$ points at the bottom, with the restrictions that the paths do not touch each other and run monotonically in the vertical direction. Two braids are considered equivalent if there exists a continuous deformation transforming one into the other. We can obtain the product of two braids by gluing the bottom of the first braid cylinder  with the top of the second braid cylinder. In this setting, a generator $\sigma_i$ is the braid in which only the strands $i$ and $i+1$ cross once, and its inverse $\sigma_i^{-1}$ is the braid in which the strands $i$ and $i+1$ cross in the opposite sense. (See Figure~\ref{fig:multi})

\bigskip

The root extraction decision problem in braid groups is the following: Given a braid $\beta \in \mathcal{B}_n$ and a number $k\in \mathbb{N}$, tell if there is $\alpha\in \mathcal{B}_n$ such that $\alpha^k=\beta$. If such an $\alpha$ exists, we call it a $k$-th root of $\beta$ and, in this case, the root extraction search problem consists in finding a $k$-th root of $\beta$. The search problem will be the one used by the proposed cryptosystems. 

When first proposed, the root extraction problem was believed to be hard to solve. The first approach was made in  1979 by Sty{\v{s}}nev \cite{Stynev1979}, where it is proved that the root extraction problem in braids is decidable, but no efficient algorithm was given to solve the search problem. Much later, in 2005, a first algorithmic attack to the problem was published in \cite{GHR} by Groch, Hofheinz and Steinwandt. In this paper the authors provide an algorithm to compute a $k$-th root of a braid, by reducing the problem to compute roots in symmetric groups. However, this algorithm does not always work. The fact that this algorithm works in most of the cases is not mathematically proven, but they perform computations over different 1000-tests simulations with different parameter choices. With most parameters, the success rate is above 90 percent. But in some tests, the algorithm fails in more than 50 percent of the cases because the computer cannot manage the quantity of roots that appear in the symmetric group. This is still left open different options to create cryptosystems based on the problem: one could use this issue with roots on the symmetric group to choose parameters that make the cryptosystem resistant to the proposed algorithm.

In 2007, Lee \cite{Lee2007} gave an algorithm that always work to extract a $k$-th root of a braid. His techniques strongly rely in an underlying algebraic structure of braid groups, called Garside structure. In his paper, he does not analyze the complexity of his algorithm, but this algorithm involves computing huge sets of braids, and there are certainly cases where the complexity of the algorithm is exponential. The most recent solution to the root extraction search problem is the one given in \cite{CGS} by the first author, Gonz\'alez-Meneses and Silvero. We say that braids have a generic property if the proportion of braids with such a property in a ball of radius $r$ in the Cayley graph tends to~$1$ as $r$ tends to infinity. Informally, a property is generic if ``most of braids'' have this property.
 In this article the problem is also approached by using Garside theory, and it is proven that generically, a $k$-th root of a braid can be computed in quadratic time. That is, the braids such that this algorithm can be applied to them are generic. More explicitly, the result is as follows: Let $\beta$ be a randomly picked braid that has a $k$-th root in a ball of radius~$r$ in the Cayley graph of $B_n$. By using standard Garside theory, this braid has an associated length. Then, the probability that we can compute a $k$-th root of $\beta$ in quadratic time (with respect to its length) tends to~$1$ as $r$ tends to infinity.  

As we will see later, this latter solution can be used to attack the cryptosystems that are completely based on the root extraction problem that have been proposed so far, having a probability of success that tends to one. This does not mean that the authentication schemes and digital signatures that we will see are not completely insecure, but that the parameters cannot be chosen randomly. That is, we know that these cryptosystems are not secure for most of the parameters, but there are some braids for which the root extraction problem cannot be solved in polynomial time. This leaves the following open question: For which braids the known algorithms to solve the root extraction problem have exponential complexity? 

\medskip

The paper is divided in the following sections: In Section~2 we describe the cryptosystems that claim to be based on the root extraction problem; in Section~3 we recall some definitions from braid theory that we need to understand the attacks; Section~4 is dedicated to the cryptanalysis of the authentication schemes and digital signature; Section~5 is for conclusions.

\subsection*{Acknowledgements}
Mar\'ia Cumplido was supported by the Spanish grants US-1263032 and P20\_01109 financed by Junta de Andaluc\'ia, and the Research Program ``Braids'' of ICERM (Providence, RI). \\ We thank Juan Gonz\'alez-Meneses for useful discussions about braid theory.

\section{Cryptosystems based on the Root Extraction Problem}

In this section, we present some protocols based on the root extraction problem, namely two authentication schemes, and a digital signature scheme.

\subsection{Authentication Schemes}

An authentication scheme is a cryptographic tool in which there is a prover~Alice who wants to convince a verifier Bob about his/her identity.
In 2005 Lal and Chaturvedi proposed two authentication schemes based on algorithmic problems related to the root extraction problem in braid groups. In 2006, another authentication scheme based on the root problem in braid groups is presented by Sibert,  Dehornoy, and Girault. In this section we present these schemes and we describe what is known about their security.

\subsubsection{Authentication Scheme I \cite{LC}}
Let $B_n$ be a braid group generated by a set of generators $\{ \sigma_1, \dots, \sigma_n \}$ with $n$ even. Write $LB_n$ for the braid group generated by $\{ \sigma_1, \dots, \sigma_{\frac{n}{2}-1} \}$ and $UB_n$ for the group generated by $\{ \sigma_{\frac{n}{2}+1}, \dots, \sigma_n \}$, so the elements of the first group commute with the elements of the second group.
\begin{description}
    \item[Phase 1: Key Generation] Alice chooses two integers $r,s \geq 2$, and two elements $a$ and $b$ in $LB_n$ and $UB_n$, respectively.
    \begin{description}
    \item[Public key] $B_n, LB_n, UB_n, X = a^rb^s, r, s$
    \item[Secret key] $a, b$
    \end{description}
    \item[Phase 2: Authentication] Bob chooses two elements $c$ and $d$ in $UB_n$ and $LB_n$, respectively, and sends to Alice $Y = c^rd^s$. Alice computes $Z = a^rYb^s$ and sends it to Bob. Finally, Bob verifies that $Z = c^rXd^s$.
\end{description}

Lal and Chaturvedi claimed that this scheme is secure because of the complexity of finding a root $x$ in a braid group when $x^m$ and $m \geq 2$ are given.  However, this scheme was attacked in \cite{Tsaban2005}, where it is proved that there is no need to extract roots to forge a signature. 

\subsubsection{Authentication Scheme II \cite{LC}}
Again, let $B_n$ be a braid group generated by a set of generators $\{ \sigma_1, \dots, \sigma_n \}$ with $n$ even. Write $LB_n$ for the braid group generated by $\{ \sigma_1, \dots, \sigma_{\frac{n}{2}-1} \}$ and $UB_n$ for the group generated by $\{ \sigma_{\frac{n}{2}+1}, \dots, \sigma_n \}$.

\begin{description}
    \item[Phase 1: Key Generation] Alice chooses two integers $r,s \geq 2$, and two elements $a\in LB_n$ and $c\in B_n$.
    \begin{description}
    \item[Public key] $X = B_n, LB_n, UB_n,a^rca^s, c, r, s$
    \item[Secret key] $a$
    \end{description}
    \item[Phase 2: Authentication:] Bob chooses an element $b\in UB_n$, and sends to Alice $Y = b^rcb^s$. Alice computes $Z = a^rYa^s$ and sends it to Bob. Finally, Bob verifies that $Z = b^rXb^s$.
\end{description}

We will describe a generic attack for this scheme in the Cryptanalysis section. This attack does not need to use the solution to the root extraction problem.

\subsubsection{Authentication Scheme III \cite{SDG}}

This scheme is based on a combination of the Conjugacy Search Problem and Root Problem in braid groups and solving the root extraction problem will suffice to obtain the secret key.

\begin{description}
    \item[Phase 1: Key Generation] Alice chooses a braid $a\in B_n$ and computes $b=a^2$.
    \begin{description}
    \item[Public key] $n,b$
    \item[Secret key] $a$
    \end{description}
    \item[Phase 2: Authentication] Repeat the following $k$ times:     \begin{itemize}
    \item Alice choses randomly a braid $r \in B_n$ and sends $x=rbr^{-1}$ to Bob.
    \item Bob sends a random bit $\epsilon$ to Alice.
\begin{itemize}
        \item If $\epsilon=0$, Alice sends $y=r$ to Bob, who then checks $x=yby^{-1}$.
         \item If $\epsilon=0$, Alice sends $y=rar^{-1}$ to Bob, who then checks $x=y^2$.
    \end{itemize}
    \end{itemize}
\end{description}

\subsection{Digital Signature}

Most of the signature schemes proposed in the braid groups are based on the difficulty to solve the conjugacy search problem. In \cite{WH}, the authors proposed a signature scheme based on the root extraction problem, and it works as follows.

\begin{description}
    \item[Phase 1: Key Generation] Alice chooses a set of $k+1$ braids $a_1,$ $a_2,\dots,a_m, \alpha\in B_n$ such that the $a_i$'s pairwise commute, and computes $b_i=\alpha a^k \alpha^{-1}$.
    \begin{description}
    \item[Public key] $n,m, b_1,\dots,b_m$
    \item[Secret key] $(a_1,\cdots,a_k)$
    \end{description}
    \item[Phase 2: Signature] Alice want to sign a message hashed into a $k$-bit binary string $h_1\dots h_r$. She randomly chooses a braid $\beta$ and computes:
$$\gamma=\beta\alpha^{-1}, \quad u=\beta \left( \prod_{i=1}^m a_i^{h_i}  \right) \beta^{-1}.  $$
Alice sends to Bob the signature $(u,\gamma)$ and Bob verifies that $$u^k=\gamma \left( \prod_{i=1}^k b_i^{h_i} \right) \gamma^{-1}.$$
    
\end{description}

\section{Theory of braids}
In this section we explain the tools that we use to attack the previous cryptosystems. Firstly, we review some basic definitions concerning the algebraic manipulation of braids.

Consider the positive monoid $B_n^+$ of $B_n$. The group $B_n$ has a very nice structure called Garside structure \cite{BrieskornSaito, DehornoyParis}. Associated to this structure, there is an element $$\Delta=(\sigma_1\dots\sigma_n-1)(\sigma_1\dot \sigma_{n-2})(\sigma_1\sigma_2)\sigma_1,$$ called the Garside element, such that $\Delta^2$ generates the center of $B_n$ if ${n>2}$. If $n=2$, $Z(B_n)=\langle \Delta \rangle$. The existence of a Garside structure implies that there is a partial order in $B_n$, $\po$, defined by $\alpha \po \beta \Leftrightarrow \alpha^{-1}\beta \in {B_n^+}$. The unique great common divisor of two braids $\alpha$ and $\beta$ with respect to this partial order is denoted by $\alpha \wedge \beta$. Also, all the elements $s$ satisfying ${1\po s \po \Delta}$ are called \emph{simple} elements.

The word problem in braids groups has been very well studied \cite{Adjan, BrieskornSaito, ElrifaiMorton, Epstein1992, DehornoyParis} and using the latter structure we can associate to each braid $\alpha$ a \emph{normal form} $\alpha=\Delta^px_1\cdots x_l$, were the $x_i$'s are simple elements and are such that $x_ix_{i+1}\wedge \Delta = x_i$ for $i=1,\dots, l-1$. The number $l$ is called the \emph{canonical length} of $\alpha$ and it is denoted by $\ell(\alpha)$.
We can also express $\alpha$ as $\alpha_1^{-1}\alpha_2$, were $\alpha_1^{-1}= 1$ and $\alpha_2=\alpha$ if $p\geq 0$, and $\alpha_1^{-1}= \Delta^p x_1 \dots x_{-p}$ and $\alpha_2=x_{-p+1}\cdots x_{l}$ otherwise. This is very useful if one wants to know if $\alpha$ lies in $LB_n$: we just need to put $\alpha$ in normal form and $\alpha\in LB_n$ if and only if $\alpha_1$ and $\alpha_2$ lie in $LB_n$. Moreover, the normal form can be computed in quadratic time with respect to the canonical length of the braid.

\medskip
Topologically, $B_n$ can also be defined as the mapping class group of the $n$-puncture disc. In this setting, we can see a generator of the braid group as the switching of two consecutive punctures. The braid group on $n$ strands acts by isometries of the curve complex of the $n$-punctured disc, whose vertices are isotopy classes of non-degenerate curves (referred just as curves). According with this action, we can use the Nielsen-Thurston classification to locate braids in three disjoint categories: a braid~$\alpha$ periodic if~$\alpha^m$ is central for some~$m$; it is reducible if it is not periodic and~$\alpha^m$ preserves some curve for some $m$; and~$\alpha$ is pseudo-Anosov if there are two measurable transverse foliations, $(\mathcal{F}^s , \mu_s )$ and $(\mathcal{F}^u , \mu_u )$, and a real
number $\lambda$ such that
$$g(\mathcal{F}^s , \mu_s ) = (\mathcal{F}, \lambda^{-1} \cdot\mu_s ),\quad g(\mathcal{F}^u , \mu_u ) = (\mathcal{F}^u , \lambda \cdot \mu_u ).$$
Being pseudo-Anosov is a generic property \cite{CarusoWiest}, so the study of their properties can be very handful when designing generic attacks. Finally, observe that a braid~$\alpha$ in~$LB_n$ has to be reducible, since the only central braids are roots of powers of~$\Delta$ \cite{CK} and~$\alpha$ fixes the curve that circle all the punctures involved in the generators of $LB_n$.

\section{Cryptanalysis}

\subsection{Authentication Scheme I}
As shown by Tsaban \cite{Tsaban2005}, one can identify as Alice in the Authentication Scheme~I without solving the root extraction problem. Just notice that to do so one just needs to recover~$a^r$ and~$b^r$. The key point to attack this scheme is that~$a^r$ and~$b^r$ lie in commuting subgroups of~$B_n$. When one computes the normal form of the braid $a^rb^r$, it is easy to recover~$a^r$ and~$b^r$. Moreover, one can compute normal forms in braid groups in polynomial time. So this is a very effective attack that always works.

\subsection{Authentication Scheme II}
Notice that if we recover~$a^s$ and~$a^r$, we can forge a signature without solving the root extraction problem. We are going to attack the following (more general) authentication scheme, with~$B_n$, $LB_n$ and $UB_n$ defined as in Section~2.1:

\begin{description}
    \item[Phase 1: Key Generation] Alice chooses two integers three elements $a_1,a_2\in LB_n$ and $c\in B_n$.
    \begin{description}
    \item[Public key] $X = a_1ca_2, c$
    \item[Secret key] $a$
    \end{description}
    \item[Phase 2: Authentication] Bob chooses two elements $b_1,b_2\in UB_n$, and sends to Alice $Y = b_1cb_2$. Alice computes $Z = a_1Ya_2$ and sends it to Bob. Finally, Bob verifies that $Z = b_1Xb_2$.
\end{description}

This cryptosystem includes Authentication Scheme~II. Our aim is to prove that using $X,Y$ and $Z$ we have a method that recovers $a_1$ and $a_2$ that generically works. First notice that $X^{-1}Z=a_2^{-1}c^{-1}Ya_2$ is conjugate to $c^{-1}Y$ by $a_2$. Hence, this scheme is based on the subgroup conjugacy search problem in braids groups, that is, to find $a_2$ we need to explore what are the elements in $LB_n$ doing the same conjugation as $a_2$. The best algorithms to solve the search conjugacy problem in braids are the ones in \cite{Gebhardt-GM, Gebhardt-GM2}, and computational experiments in \cite{Gebhardt2005} showed that the algorithms efficiently provide solutions that are computable in most of the cases. The conjugacy search problem has been solved in \cite{KLT} for every Garside group and, in particular, for every braid group. The algorithm is a modification of the solution of the conjugacy problem and the probabilistic efficiency is the same.  

We are going to describe another method to attack this problem. This also relies on the solution of the general conjugacy problem and it does not work for every case. However, it generically works and it is easy to describe if one wants to use the (already programmed) solution to the general conjugacy problem as a blackbox.  We can use the algorithms in \cite{Gebhardt-GM, Gebhardt-GM2} to compute $\alpha$ doing the same conjugation as $a_2$. We want to recover $a_2$ from $\alpha$. Notice any element $\alpha$ that conjugates $X^{-1}Z$ to $c^{-1}Y$ has the form $\alpha= a_2 z$, where $z$ is an element in the centralizer of $X^{-1}Z$. In \cite{GM-Valladares} it is proven that, generically, the centralizer of a braid is generated by two mutually commuting elements $v=\Delta^{e}$ and $w$, that can be computed in polynomial time. Then we can write $\alpha=v^rw^sa_2$ and we can recursively multiply $\alpha$ on the left by $w^{-1}$ until we reach and element of the form $\Delta^m a'$, where $a'\in LB_n$ (this can be easily checked by computing the normal form of the element). The last step to see that this algorithm is efficient is to show that there is an upper bound for $s$. Generically, the canonical length of $w$ linearly increases with respect $s$; this is due to the generic rigidity proved in \cite{CarusoWiest}. Also notice that $\ell(w^s)$ is bounded by  $\ell(\alpha)+\ell(a_2)$. The canonical length of $a_2$ is upper bounded by the parameters of the cryptosystem and, generically, by \cite{Gebhardt2005,GM-Valladares} the canonical length of $\alpha$ (and any conjugacy element from $X^{-1}Z$ to $c^{-1}Y$) linearly depends on the maximal canonical length of the parameters. 

We have proven that we can obtain and element $a'$ doing the same conjugacy as $a_2$ using either the algorithm in \cite{KLT} or the previously describe method, but we want to find precisely $a_2$. To see that the scheme is not safe, we prove that generically the only conjugacy element from $X^{-1}Z$ to $c^{-1}Y$ that is contained in $LB_n$ is $a_2$. We know by \cite{CarusoWiest} that $X^{-1}Z$ is generically pseudo-Anosov and it is well known that elements in the centralizer of a pseudo-Anosov braid $\beta$ cannot be reducible, as this would imply that $\beta$ is reducible. Suppose that $\alpha\in LB_n$, hence $v^rw^s\in LB_n$ cannot be pseudo-Anosov and it must be trivial, because the only power of $\Delta$ that lies in $LB_n$ is $\Delta^0=1$.  We have then proven that $a'=a_2$. We can do the same process to obtain $a_1$. The previous discussion means that Algorithms~\ref{algorithm1} and \ref{algorithm2} will efficiently forge Alice's signature in almost every case.

\begin{algorithm}
\caption{First algorithm to forge the signature of Authentication Scheme~II}\label{algorithm1}

\SetKwInOut{Input}{Input}\SetKwInOut{Output}{Output}

\BlankLine

\textbf{Input:} {$X,Y,Z,c$} \\  \textbf{Output:} {$a_1,a_2$}

\BlankLine
\BlankLine

$a:=$ conjugacy element from $X^{-1}Z$ to $c^{-1}Y$ that lies in $LB_n$ (use algorithm in \cite{KLT});

$b:=$ conjugacy element from $ZX^{-1}$ to $Y$ that lies in $LB_n$ (use algorithm in \cite{KLT});

\Return{$b^{-1},a$}

\end{algorithm}

\begin{algorithm}
\caption{Second algorithm to forge the signature of Authentication Scheme~II}\label{algorithm2}

\SetKwInOut{Input}{Input}\SetKwInOut{Output}{Output}

\BlankLine

\KwData{$X,Y,Z,c$}
\KwResult{$a_1,a_2$}

\BlankLine
\BlankLine

$\alpha:=$ conjugacy element from $X^{-1}Z$ to $c^{-1}Y$ (use algorithm in \cite{Gebhardt-GM2});

$\{\Delta^e,w\}:=$ generators of the centralizer of $X^{-1}Z$ (use algorithm in \cite{GM-Valladares});

$a \gets \alpha$;

\While{$a\neq \Delta^m a'$, for some $m\in \mathbb{Z}$ and $a'\in LB_n$}{
$a\gets w^{-1}a$;
}

$\alpha:=$ conjugacy element from $ZX^{-1}$ to $Y$ (use algorithm in \cite{Gebhardt-GM2});

$\{\Delta^e,w\}:=$ generators of the centralizer of $X^{-1}Z$ (use algorithm in \cite{GM-Valladares});

$b \gets \alpha$;

\While{$b\neq \Delta^m b'$, for some $m\in \mathbb{Z}$ and $b'\in LB_n$}{
$b\gets w^{-1}b$;
}

\Return{$b'^{-1},a'$}

\end{algorithm}

\bigskip

\subsection{Authentication Scheme III}

The authors of Authentication Scheme III make the remark that the public key should be picked so that the square root extraction problem will be hard to solved. However, this was made when no effective generic algorithm existed. Nowadays, if the public key is picked randomly, the secret key can be obtain generically very fast by using the method in \cite{CGS}. Also, there is no method to generate braids for which the algorithm in \cite{CGS} does not work.  

\subsection{Digital Signature}
The authors of the proposed digital signature in \cite{WH} proved that one can forge a signature if one can solve the root extraction problem. Indeed, if we compute $v= \prod_{i=1}^k b_i^{h_i}$ and we extract a $k$-th root $w$ of $v$ and choose a random braid $\beta'$, then $(\beta'w\beta'^{-1},\beta')$ is an accepted signature: just check that $(\beta'w\beta'^{-1})^k=\beta'w^k\beta'^{-1}= \beta'v\beta'^{-1}$.

Then, the algorithm in \cite{CGS} generically works to forge a signature.

\section{Conclusions}
The Authentication Scheme~I was already efficiently broken. The Authentication scheme~II does not really depends on the solution of the $k$-th root extraction problem, but in the solution of the subgroup conjugacy problem in braids. We have described a generic attack to identify as Alice using the algorithm in \cite{KLT}. We can forge the signature in the Authentication Scheme~III and the Digital Signature by using the solution to the root extraction problem, so we can use the algorithm in \cite{CGS} to attack those cryptosystems.

This means that the last three proposed cryptosystems are not secure if we randomly choose the parameters. However, generic attacks do not work in every case.  It is an open question whether we can build a method to generate braids such that solving the root extraction problem or the subgroup conjugacy search problem takes exponential time.

\bibliographystyle{plain}
\bibliography{bib}

\end{document}